\documentclass[aps,prd,twocolumn,showpacs,groupedaddress]{revtex4-1}

\pdfoutput=1
\usepackage{amssymb}
\usepackage{amsmath}
\usepackage{hyperref}

\begin{document}


\title{A variational principle for  asymptotically Randall-Sundrum black holes}


 
 \author{Scott Fraser}
\email[Email address: ]{scfraser@calpoly.edu}
\affiliation{Department of Physics, 
California Polytechnic State University,  
San Luis Obispo, California 93407}

\author{Douglas M. Eardley}
\email[Email address: ]{doug@kitp.ucsb.edu}
\affiliation{Department of Physics, University of California, Santa Barbara, California 93106}



\begin{abstract}
We   prove
the following  variational principle for asymptotically Randall-Sundrum (RS)
black holes,  
  based on  the first law of black hole mechanics:
 Instantaneously static initial data that extremizes
the mass yields a static black hole,
for variations  at 
 fixed     apparent horizon area, AdS
curvature length, cosmological constant,  brane tensions, and  RS brane warp factors.  
This variational principle is valid  with either  two branes (RS1) or  one brane (RS2), and is  applicable to  variational  trial       solutions.
\end{abstract}



\maketitle

\section{Introduction \label{intro}}

The first law   for a static or stationary black hole  
expresses  conservation of energy,
 and takes the  general form
 $\delta M = (\kappa/8\pi G) \delta A + \sum_i p_i \,\delta Q_i$.  
This relates the variations of mass $M$, horizon area $A$,
and other physical quantities $Q_i$. Thus, if a  black hole is  static or stationary,    it   
extremizes  $M$  under variations  that hold  constant the
remaining variables  ($A$, $Q_i$). 
The  converse        of this statement
motivates  a   variational principle:
{\it If a black hole's spatial geometry  is  initially   static (or    initially stationary) and
extremizes the mass  with other physical variables held fixed, then  
the black hole
is static  (or stationary).} 

In this variational principle, the  specific  variables  to   hold fixed  depend on    
the    first law. The  appropriate area to hold fixed is that of the  apparent horizon, which 
  is   determined   by the  
  spatial geometry alone (unlike   the   event horizon, which is a global  spacetime property).
The apparent horizon
 generally lies inside  the event horizon, and  coincides with it
  for a static or stationary spacetime.

A version of the above   variational principle    was proved
 by Hawking for stationary black holes \cite{hawking}, and was extended to Einstein-Yang-Mills theory 
  by Sudarsky and Wald   
 \cite{sudarsky-wald, *chrusciel-wald}.
 These results are for asymptotically flat black holes in four spacetime dimensions.  There is
 also much ongoing interest in  extra dimensions, including the Randall-Sundrum (RS) 
 braneworld models    \cite{RS1, RS2}.  
In this  paper, we prove a  
 version of the above variational principle  for  asymptotically RS  black holes.

The RS models   are phenomenologically interesting, and  have holographic interpretations \cite{RS-AdS-CFT} in   the AdS/CFT  correspondence.
In the RS models, our observed universe is
a brane surrounded by an AdS bulk. The bulk is warped by  a negative  
cosmological constant.
The RS1 model \cite{RS1}  has two branes of opposite tension, with
our universe  on the negative-tension brane.  
 Tuning the interbrane
distance   appropriately predicts the 
   production of small black holes at  TeV-scale collider energies 
    \cite{Banks-Fischler, *Giddings}, and   LHC experiments  
  \cite{LHC-RS-extra-dimensions, *LHC-RS-black-holes} continue to test this hypothesis. 
  
 In the RS2 model \cite{RS2}, 
our  universe resides on  the positive-tension brane,
with the negative-tension brane  removed to infinite distance.
Perturbations of RS2   reproduce
 Newtonian gravity  at large distance  on the brane, while in
 RS1 this requires a mechanism to stabilize the interbrane distance \cite{Tanaka}.
In  RS2,   solutions for static black holes    on the brane   have been found  numerically,   for both
    small black holes
\cite{Kudoh-smallBH-1, *Kudoh-smallBH-2, *Kudoh-smallBH-6D} and large  
 black holes  \cite{Figueras-Wiseman, *Abdolrahimi}, compared to the AdS curvature length.
The only known exact  analytic black hole
solutions
are the static and stationary  solutions   \cite{ehm, *ehm-2} in a lower-dimensional version of RS2.

In    \cite{paper-1-first-law}, we proved a  
first law     for a static
asymptotically
RS black hole, including variations of
the    AdS curvature length, cosmological constant, brane tensions, and RS brane warp factors.
This first law   motivates the   following
 variational principle,   which we   prove in this paper.

 {\bf Variational principle for  asymptotically  RS black holes:} 
  {\it 
Instantaneously   static   initial data that 
extremizes the mass    $M$ is  initial data for
a static    black hole, for   variations
  that leave fixed the apparent horizon area $A$, 
  the AdS curvature length $\ell$, cosmological constant $\Lambda$, brane tensions  $\lambda_i$, 
and RS values (at spatial infinity) of the warp factors   $\Omega_i$ on each brane.}

This paper is organized as follows.  We review  the RS spacetimes in section \ref{RS-review}, 
and   prove the variational principle in section \ref{vp-proof}.
 We demonstrate an explicit  application of the  variational principle   in section \ref{black string application},
and conclude in section \ref{conclusion}.
Throughout this paper, we   use two branes, so  our results
 apply to 
     either   RS1   or   RS2  
in the appropriate limit.
We work on the orbifold  region  (between the branes) and  use  $D$ spacetime dimensions. 
A timelike surface has metric $\gamma_{ab}$, extrinsic curvature
$K_{ab}=\gamma_a{}^c\nabla_c n_b$, and outward unit normal     $n_b$.
A spatial hypersurface   $\Sigma$
has unit normal $u_a$, metric $h_{ab}$, and covariant derivative $D_a$.
Each  boundary $B$ of $\Sigma$
  has metric  $\sigma_{ab}$, extrinsic curvature $k_{ab}=h_a{}^c D_c n_b$,
   and outward unit normal     $n_b$.
The  boundaries $B$ of $\Sigma$ are:
 $B_1$, $B_2$ (the branes), 
 $B_\infty$ (spatial infinity), and $B_{H}$
(the apparent horizon).

\section{The Randall-Sundrum spacetimes
\label{RS-review}}

The RS spacetimes  \cite{RS1,RS2} are portions of 
 an anti-de Sitter  (AdS) spacetime, with metric
\begin{equation}
\label{g_RS}
ds^2_{\rm RS}
= 
\Omega(Z)^2 \left(- dt^2 + d\rho^2 + \rho^2d\omega_{D-3}^2 + dZ^2 \right) \ .
\end{equation}
Here $d\omega_{D-3}^2$ denotes the unit $(D-3)$-sphere.  
The    warp factor is   $\Omega(Z)=\ell/Z$,  with values   $\Omega_i$ on each brane.
Here $\ell$ is the AdS curvature length, related to the bulk cosmological constant $\Lambda<0$    given below.
The RS1 model \cite{RS1} contains two  branes, which are the    surfaces      $Z=Z_i$ 
with brane  tensions  $\lambda_i$, where $i=1,2$. 
The brane tensions $\lambda_i$ and      bulk cosmological constant  $\Lambda$ are  
  \begin{equation}
  \label{RS-tensions}
\lambda_1 =  -\lambda_2    =
\frac{2(D-2)}{8\pi G_D\ell}
\ , \quad
\Lambda    =   -\,\frac{(D-1)(D-2)}{2\ell^2} \ .
\end{equation} 
The dimension $Z$ is   compactified on the orbifold $S^1/\mathbb{Z}_2$ and the 
  branes have    orbifold mirror
 symmetry:
in the covering space,  symmetric points across a brane are identified.  
There is a  discontinuity   in the extrinsic curvature $K_{ab}$ across each brane
 given by the Israel condition  \cite{israel}. Using
 orbifold symmetry, the Israel condition
 requires  the   extrinsic curvature at each brane to satisfy
\begin{equation}
\label{RS-K}
2 K_{ab}
 = \frac{\varepsilon}{\ell}\gamma_{ab}
\ , \quad
2k_{ab} = \frac{\varepsilon}{\ell}  \sigma_{ab} \ ,
\end{equation}
where $\varepsilon=\pm 1$   is the sign of each brane tension.
The RS2 spacetime \cite{RS2}
is   obtained   from RS1
 by   removing the  negative-tension brane (now   a regulator)
 to infinite distance ($Z_2 \rightarrow \infty$)
and the orbifold  region has $Z \ge Z_1$.

\section{Proof of the variational principle \label{vp-proof}}

 Our main step, which we carry out in section \ref{main step},
  reduces the proof to   
  two auxiliary boundary value problems, which are the topics of
  section \ref{BVPs}.
  
  Our setup is general: it applies to a black hole  localized on   a brane,   or    isolated    in the bulk (away from   either brane),  and also applies to the asymptotically RS black string \cite{black-string}.
Our  key assumptions  will be the following.  First, we   assume our
     initial data  $h_{ab}$   is instantaneously static.  We will also assume  
the variations  $\delta h_{ab}$          extremize the mass ($\delta M=0$), while
 holding fixed      the apparent horizon area $A$ and   the quantities
   ($\ell$,   $\Lambda$,    $\lambda_i$, $\Omega_i$).

\subsection{Main  proof \label{main step}}
 
 Our method   closely follows our proof of the first law for
static asymptotically RS black holes \cite{paper-1-first-law},  which is based on the
Hamiltonian formulation of general relativity.
The full
Hamiltonian contains a bulk term
and surface terms.  The bulk  term is   defined
 on
an initial data surface $\Sigma$, 
\begin{equation}
\label{H_sigma} H_{\Sigma} =\int_{\Sigma} d^{D-1} x\, \left( N
{\cal C}_0 + N^a {\cal C}_{a} \right) \ .
\end{equation}
Here $N$ and $N^a$ are
the   lapse and shift functions
 in the standard decomposition of the     metric.   
  Our focus is the initial data ($h_{ab}$, $p^{ab}$) where  $h_{ab}$  is the spatial metric 
 and
  $p^{ab}$ is its  canonically conjugate momentum,
\begin{equation}
   16\pi G_D \, p^{ab}=\sqrt{h} {\cal K}^{ab} - {\cal K}\, h^{ab} 
   \quad , \quad
    {\cal K}_{ab} = h_a{}^c \nabla_c u_b
    \ .
\end{equation}
Initial data must satisfy     constraints, 
${\cal C}_0 = 0$ and ${\cal C}_a= 0$, which we henceforth assume, where
\begin{eqnarray}
 \nonumber {\cal C}_0 &=&
\frac{\sqrt{h}\left(2\Lambda - {\cal R}\right)}{16\pi G_D}  +
\frac{16\pi G_D}{\sqrt{h}} \left( p^{ab} p_{ab} - \frac{ p^2}{D-2}
\right) \ ,
\\*
 {\cal C}_a &=& -2 D_b p_a{}^b \ .
\end{eqnarray}
Here
$\cal R$ and $D_a$ are the Ricci scalar and
  covariant derivative
associated 
with $h_{ab}$.
We now consider the change  $\delta  H_{\Sigma}$ under
 variations  ($\delta h_{ab}$, $\delta p^{ab}$). 
One finds   
 $\delta{\cal C}_0$ and $\delta{\cal C}_a$
involve derivatives ($D_c\delta h_{ab}$, $D_c\delta p^{ab}$).
Integrating  by parts to remove these
derivatives yields
  surface terms  $I_B$,  
\begin{equation}
  \delta H_{\Sigma} =  \int_{\Sigma} d^{D-1}x
 \left(
 {\cal P}^{ab}\delta h_{ab} + {\cal H}_{ab}\delta p^{ab}
 \right)
+
\sum_B I_B
\label{delta_H_sigma-1}
\ .
\end{equation}
The quantities  (${\cal P}^{ab}$, ${\cal H}_{ab}$) appear in the   time evolution equations,
which involve the
 Lie derivative (denoted by an overdot)
along the vector  $t^a= N u^a + N^a$,
 \begin{equation}
 \label{evolution}
 \dot h_{ab} = {\cal H}_{ab} \ , \quad
 \dot p^{ab}=-{\cal P}^{ab} \ .
\end{equation}
We will not need the 
most general forms of  ${\cal P}_{ab}$,   ${\cal H}_{ab}$, and  $I_B$.
Will give their  simplified forms below, after implementing some of our key assumptions.
We now assume our variations take
  one solution of the constraints 
 to another solution of the constraints, so we take
  $\delta{\cal C}_0=0$ and $\delta{\cal C}_a=0$.
Then the variation of (\ref{H_sigma})
immediately
  gives     
\begin{equation}
\label{delta_H_sigma-2}
\delta  H_{\Sigma}=0 \ .
\end{equation}
We henceforth assume the initial data  is instantaneously static,  for which
  $p^{ab}=\delta p^{ab}=0$
  and we take $N^a=0$.  
  One then finds
 ${\cal H}_{ab}=0$, so   (\ref{delta_H_sigma-1}) and   (\ref{delta_H_sigma-2}) give
\begin{equation}
\label{vp-delta_H_sigma}
\int_{\Sigma} d^{D-1}x \,{\cal P}^{ab}\delta h_{ab}
+
\sum_B I_B
=0 \ ,
\end{equation}
In what follows, (\ref{vp-delta_H_sigma})
  will be our primary equation, 
where
\begin{equation}
\label{vp-Pab-initial}
 {\cal P}^{ab}   =   \frac{\sqrt{h}}{16\pi G_D}
  \left(
{\cal R}^{ab}+ h^{ab}D_c D^c  - D^a D^b
  \right)N \ .
\end{equation}
For   instantaneously static initial data, 
the   constraint ${\cal C}_0=0$  simplifies to
  ${\cal R}=2\Lambda$
and $\delta {\cal C}_a$ vanishes identically. The  
   linearized constraint ($\delta{\cal C}_0=0$) 
simplifies to
\begin{equation}
\label{vp-linearized-constraint-1-red}
 \left({\cal R}^{ab} + h^{ab}D^c D_c -D^a D^b\right)
 \delta h_{ab}  = 0 \ .
\end{equation}

We now evaluate the   terms $I_B$ in (\ref{vp-delta_H_sigma}).
We found the values of    $I_{B_i}$   (at each brane) and 
$I_{B_\infty}$ (at spatial infinity)   in  \cite{paper-1-first-law}, including the
variations of    quantities ($\ell$, $\Lambda$, $\lambda_i$, $\Omega_i$)   
  held  constant here  by  assumption. In this case,
 \cite{paper-1-first-law} gives
\begin{equation}
\label{surface-terms-values}
I_{B_i} = 0  \ , \quad
I_{B_\infty}= -\delta M    \ .
\end{equation}
Additionally, we have $\delta M=0$, by    our   assumption  
 of a mass extremum, so $I_{B_\infty}=0$.
At the  apparent
 horizon,
\begin{equation}
\label{surface-terms}
 I_{B_H} =  \int d^{D-2}x  \sqrt{\sigma} 
{\cal A}^{bcd}  \left[(D_bN)\delta h_{cd}
- ND_b\delta h_{cd} \right]  \ ,
\end{equation}
where 
\begin{equation}
16\pi G_D \, {\cal A}^{bcd}= n_a(h^{ac}h^{bd} -h^{ab}h^{cd}) \ .
\end{equation}
The boundary condition  on the lapse is $N=0$, whence
$D_aN = -f n_a$, where
$f^2 =  (D^bN)(D_bN)$.
Then     
  (\ref{surface-terms}) is
\begin{equation}
 I_{B_H}=
 \frac{1}{8\pi G_D} \int  d^{D-2}x \, f\,
\delta\sqrt{\sigma} \ ,
\end{equation}
using 
$\sigma_{ab}=h_{ab}- n_a n_b$
and 
$
 \delta \sqrt{\sigma}
=  \sqrt{\sigma}\sigma^{ab}\delta \sigma_{ab}/2
$.
For convenience, we  now  choose to set $I_{B_H}=0$
  using the following gauge transformation,
\begin{equation}
\label{horizon gauge}
\delta\sigma_{ab}  \rightarrow  \delta\sigma_{ab}   +   2{\cal D}_{(a}\xi_{b)} \ ,
\quad
\sigma^{ab}\delta \sigma_{ab}
   \rightarrow
 0  \ ,
\end{equation}
where ${\cal D}_a$ is the covariant derivative associated
 with $\sigma_{ab}$.
If we let $\xi_a={\cal D}_aF$, then 
$
\sigma^{ab}\delta \sigma_{ab}
   \rightarrow
 0
$
requires
\begin{equation}
\label{vp-throat-gauge}
-\sqrt{\sigma}\,{\cal D}^a{\cal D}_a F = \delta \sqrt{\sigma} \ .
\end{equation}
Note the apparent horizon is a closed
surface (which is most clearly seen in the  covering space, if the apparent horizon  intersects a brane).
A solution $F$ to (\ref{vp-throat-gauge}) on a closed surface is well known to exist
if  and only if the integral of the right-hand side of  (\ref{vp-throat-gauge})  vanishes.  This  integral 
  is   simply   $\delta A$, which indeed vanishes    since we   hold  $A$ constant.  Thus     
  a solution   $\xi_a$ exists to achieve (\ref{horizon gauge}), and we henceforth set $I_{B_H}=0$.
Since  $I_{B_i}=I_{B_\infty}=0$ and we      set $I_{B_H}=0$,
  our primary equation (\ref{vp-delta_H_sigma})    simplifies to
\begin{equation}
\label{vp-no-go-converse-1}
\int_{\Sigma} d^{D-1}x  \, {\cal P}^{ab}\delta h_{ab} =0
  \ .
\end{equation}

Our goal is   to conclude   that  the
initial  geometry $h_{ab}$
evolves to a static spacetime. The well known condition for this is 
 that ${\cal P}^{ab}=0$
on   $\Sigma$.  
We cannot, however, immediately conclude that  ${\cal P}^{ab}=0$  from (\ref{vp-no-go-converse-1}),
 because not all of the variations $\delta h_{ab}$
are arbitrary: the
linearized constraint  (\ref{vp-linearized-constraint-1-red})   removes one
degree of freedom, which  can be taken
  as       $h^{ab}\delta h_{ab}$  or as
  the variation $\delta h$ of the
 determinant $h$. These    are related by
$
h^{ab} \delta h_{ab}
=
 \delta h/h$.  
 As an identity, we may  decompose $\delta h_{ab}$ into a trace-free ($TF$)
part and a part proportional to $\delta h$:
\begin{equation}
\label{variations-decomp}
 \delta h_{ab} =
(\delta h_{ab})^{TF}
  +
\frac{1}{D-1}\left(\frac{\delta h}{h}\right)h_{ab} \ .
\end{equation}
Using   (\ref{variations-decomp}),
our primary equation (\ref{vp-no-go-converse-1})  then  becomes
\begin{equation}
\label{vp-no-go-converse-2}
 \int_{\Sigma} d^{D-1}x
 \left[ ({\cal P}^{ab})^{TF}(\delta h_{ab})^{TF}
  +
 \frac{{\cal P}\, \delta h}{(D-1)h}
  \right]
  =0 \ ,
\end{equation}
where   
\begin{eqnarray}
\label{P-decomp}
  {\cal P}_{ab} &=&
( {\cal P}_{ab})^{TF}
  +
\frac{\cal P}{D-1} h_{ab}  \ ,
\\
{\cal P} &=& h_{ab}{\cal P}^{ab} \ .
\end{eqnarray} 
The arbitrary  variations are $(\delta h_{ab})^{TF}$,  
subject to  
  smoothness at the apparent horizon,
  $(\delta h_{ab})^{TF} \rightarrow 0$ at $B_\infty$,
  and       boundary conditions at the branes that we will specify 
   in the next section.
 As a completeness check,  the  arbitrary  variations  $(\delta h_{ab})^{TF}$
   alone should 
determine the dependent quantity  $\delta h$, which
we   verify below by showing the
linearized constraint   (\ref{vp-linearized-constraint-1-red}) 
is a well posed boundary value problem  for $\delta h$.

Our proof   then  reduces to 
showing     
     ${\cal P}=0$, which  allows 
us to conclude  
from (\ref{vp-no-go-converse-2}) that
 $({\cal P}^{ab})^{TF}=0$, since
  $(\delta h_{ab})^{TF}$
are   arbitrary variations.
It   then follows  from (\ref{P-decomp}) that
  ${\cal P}^{ab}=0$, which is the desired result.
The statement  ${\cal P}=0$ 
is a  boundary value problem for $N$ that we demonstrate   
 is solvable in the following section,   which
completes our proof of the variational principle.

\subsection{Auxiliary boundary value problems \label{BVPs}}

The  boundary value problem for  $N$     is 
   \begin{subequations} \label{N-BVP}
\begin{eqnarray}
\label{vp-N-eom}
 D^a D_a N  -    \frac{(D-1)}{\ell^2}  N
 &=& 0 \ ,
 \\*
\label{vp-N-bc-brane}
\left(n^a D_a   N   -      \frac{\varepsilon}{\ell}\, N \right)\Big|_{B_i}
  &=&   0  \ ,
  \\*
  \label{vp-N-bc-throat}
    N \, \Big|_{B_H} &=&   0  \ ,
    \\*
    \label{vp-N-bc-infty}
    N\, \Big|_{B_\infty} &\rightarrow& \Omega  \ .
\end{eqnarray}
   \end{subequations}
Here,   $\Omega$ is the warp factor of the asymptotic RS solution (\ref{g_RS}) and
$\varepsilon=\pm 1$ is the sign of each brane tension.
The result (\ref{vp-N-eom})  follows from  setting
  ${\cal P}= h_{ab}{\cal P}^{ab}=0$,
  using
(\ref{vp-Pab-initial}) and
    ${\cal R}=2\Lambda$.
The boundary conditions (\ref{vp-N-bc-throat}) and   (\ref{vp-N-bc-infty})
are straightforward.
Our main concern is the brane boundary condition   (\ref{vp-N-bc-brane}), which
 results from  using
\begin{equation}
2K_{ab} = n^c \partial_c \gamma_{ab} + \gamma_{ac} \partial_b n^c +  \gamma_{bc} \partial_a n^c \ .
\end{equation}
Now    $\gamma_{tt}=-N^2$ and $\gamma_{ta}=0$  gives
$
2K_{tt} = n^c \partial_c (-N^2) 
$,
and the Israel condition $K_{tt} = (\varepsilon/\ell)\gamma_{tt}$ then   gives
 (\ref{vp-N-bc-brane}).

 As shown in  \cite{inverse-positivity}, the following approach
  can    put a Robin boundary condition (\ref{vp-N-bc-brane})   in a standard   form while keeping
    its associated elliptic equation  (\ref{vp-N-eom})  in a   divergence form.
Let $v_a$ be any vector field and  define   ${\cal V}_aN = (D_a  - v_a)N$.
Then  (\ref{vp-N-eom}) and  (\ref{vp-N-bc-brane})  
become
  \begin{equation}
  \label{vp-N-eom-2}
 D^a {\cal V}_a  N 
 + v^a D_a N
+  \left[    D_a v^a -  \frac{(D-1)}{\ell^2}\right] N
=0 
\end{equation}
and
\begin{equation}
\label{vp-N-bc-brane-2}
\left[n^a {\cal V}_a N+ \left(n^a v_a  -  \frac{\varepsilon}{\ell}\right)N \right]\Big|_{B_i}
=   0 \ .  
\end{equation}
As in   \cite{inverse-positivity}, we now
 choose $v_a$ so      $(n^av_a - \varepsilon/\ell) \ge 0$ at     $B_i$,
which is the usual prerequisite for applying an existence theorem  to a    boundary 
   value problem of the form
    (\ref{vp-N-eom-2})\textendash(\ref{vp-N-bc-brane-2}).
For example, we choose $v_a = - \widetilde n_a/\ell$,
where $\widetilde n_a$  is 
any  vector field,
  pointing from   
   $B_1$ to   $B_2$,
    that   interpolates
    from the inward unit normal  ($-n_a$) of  $B_1$
to the
outward unit normal $n_a$  of $B_2$. 
Then $n^a\widetilde n_a = -\varepsilon$ at each brane $B_i$,
and
$v_a = - \widetilde n_a/\ell$     gives $(n^av_a - \varepsilon/\ell) = 0$ in  (\ref{vp-N-bc-brane-2}).
With the brane  boundary conditions in  standard form,  and   the remaining
 standard (Dirichlet) boundary conditions, 
  (\ref{vp-N-bc-throat}) and   (\ref{vp-N-bc-infty}),
 we   then  readily infer that the  boundary value problem
(\ref{N-BVP}) for $N$ is solvable.

We now turn to the boundary value problem for
  $\delta h$, 
which  we will state in terms of a scalar quantity $(\delta h/h)$,
   \begin{subequations} \label{dh BVP}
\begin{eqnarray}
\label{vp-dh-eom}
 D^a D_a (\delta h/h)   -  \frac{(D-1)}{\ell^2}  \, (\delta h/h)
 &=& f_{\Sigma} \ ,
 \\*
\label{vp-dh-bc-brane}
\left[n^a D_a   (\delta h/h)   -    \frac{\varepsilon}{\ell} \, (\delta h/h)
\right]\Big|_{B_i}
  &=&   f_{i} \ ,
  \\*
\label{vp-dh-bc-throat}
n^a D_a (\delta h/h)   \Big|_{B_H} &=&   f_H \ ,
    \\*
 \label{vp-dh-bc-infty}
    (\delta h/h)\ \Big|_{B_\infty} &\rightarrow& 0 \ .
\end{eqnarray}
   \end{subequations}
   As above,  $\varepsilon =\pm 1$ is the sign of each brane tension.
The  result   (\ref{vp-dh-eom})   follows from
 substituting  (\ref{variations-decomp})
 into the  linearized
constraint 
(\ref{vp-linearized-constraint-1-red})
 with ${\cal R}=2\Lambda$.
We will give the source terms and
  derive the boundary conditions below.

The key point  is that
(\ref{dh BVP}) is a well posed 
boundary value problem. The  
 elliptic equation 
  (\ref{vp-dh-eom})   and the   boundary conditions 
   (\ref{vp-dh-bc-brane})  
 are  similar in  form to 
   (\ref{vp-N-eom}) and  (\ref{vp-N-bc-brane})  in the  previous  boundary value 
   problem (\ref{N-BVP}). The remaining boundary conditions,
 (\ref{vp-dh-bc-throat}) and (\ref{vp-dh-bc-infty}), are   well known types:
 Neumann and  Dirichlet, respectively.

In the remainder of this section, we
provide the details of the source terms and
  the boundary conditions in (\ref{dh BVP}).
The source terms in (\ref{dh BVP}) are  
\begin{eqnarray*}
   f_{\Sigma} &=&
  \frac{D-1}{D-2} \left[D^a D^b(\delta h_{ab})^{TF}
   -
  {\cal R}^{ab} (\delta h_{ab})^{TF} 
  \right]\ ,
\\*
 -f_{i} &=&  \frac{D-1}{D-2} 
 \left[
\sigma^{ab} n^c D_c (\delta
h_{ab})^{TF} +  \frac{\varepsilon D}{\ell}n^a n^b (\delta h_{ab})^{TF}
 \right]
 \ ,
\\*
  f_H &=&
  \frac{1}{D-2}
  \left[
2k^{ab} (\delta h_{ab})^{TF} - \sigma^{ab} n^c D_c (\delta h_{ab})^{TF}
\right] \ .
\end{eqnarray*}
 The boundary conditions on $\delta h$  are given
  by
varying those  on $h_{ab}$, which
at the apparent horizon and the branes 
 involve the extrinsic curvature $k_{ab}=\sigma_a{}^c D_c n_b$,
\begin{equation}
\label{k-boundary-conditions}
k \Big|_{B_H} =   0
\ , \quad
k_{ab} \Big|_{B_i} = \frac{\varepsilon}{\ell} \, \sigma_{ab} \ .
\end{equation}
By varying these, we obtain
\begin{equation}
\label{unconstrained-var}
\delta k \Big|_{B_H} =   0
\ , \quad
\delta k \Big|_{B_i} =   0
\ , \quad
\delta k_{ab} \Big|_{B_i} = \frac{\varepsilon}{\ell} \, \delta \sigma_{ab} \ .
\end{equation}
 To evaluate these,
we use the   general results
   \begin{subequations}
\begin{eqnarray}
 \label{unconstrained-var-dk_ab}
2\delta k_{ab}  &=&
 \left(n^c n^d\delta h_{cd}\right)k_{ab}
 -
 \sigma_a{}^c \sigma_b{}^d  n^f
J_{cdf} \ ,
\\
\label{unconstrained-var-dk}
-2\delta k  &=&  2k^{ab} \delta \sigma_{ab}  -  k\,n^a n^b \delta h_{ab}
+   \sigma^{ab} n^c J_{abc} \ ,
\\
J_{abc} &=&  D_{a}\delta h_{bc} + D_{b}\delta h_{ac} - D_c\delta h_{ab} \ .
\end{eqnarray}
   \end{subequations}
The boundary conditions at the branes  (\ref{vp-dh-bc-brane})  
and the apparent horizon   (\ref{vp-dh-bc-throat}) 
result  from
evaluating  $\delta k=0$ using
 (\ref{variations-decomp}),
 (\ref{k-boundary-conditions}),
 and
 (\ref{unconstrained-var-dk}).
The last relation in
 (\ref{unconstrained-var})
expresses
brane  boundary conditions for $(\delta h_{ab})^{TF}$, since
it
reduces
to a form   independent of $\delta h$
after
substituting
(\ref{RS-K}),
(\ref{variations-decomp}),
  (\ref{vp-dh-bc-brane}), and (\ref{unconstrained-var-dk_ab}).

\section{A direct application of the variational principle \label{black string application}}

Here we    
  demonstrate  the utility of the variational principle, 
  by applying it   to a trial  solution
and reproducing
   the static asymptotically   RS    black string        \cite{black-string},
which is  the only  known exact     solution for   an
  asymptotically   RS  
  black  object  in 5-dimensional spacetime. 
We first  specify  a  trial  geometry for an initially static a black string. After evaluating
the apparent horizon area $A$ and mass $M$, we   
then   apply the variational principle.

A   black string  
is a set of lower dimensional black holes   stacked in an extra dimension $Z$,
which is how we    will   construct the trial geometry.  We take
\begin{equation}
\label{ansatz}
ds^2 = 
\Omega(Z)^2
\left[
\Psi^4({\bf x},Z) \, d{\bf x}^2+ dZ^2
\right] \ ,
\end{equation}
where $\Omega=\ell/Z$ and the branes are  the surfaces
  $Z=Z_1$ and $Z=Z_2$.
We  will  take
\begin{equation}
\label{Psi-1}
\Psi = 1 + \frac{\rho_0}{2}\left(
\frac{1}{|{\bf x}+{\bf x}_0|} +  \frac{1}{|{\bf x}-{\bf x}_0|}
\right) \ .
 \end{equation}
Here   $\rho_0 >0 $  is a constant and
 ${\bf x}_0= (0,0, d)$, using
  Cartesian coordinates ${\bf x} =(x, y, z)$ 
with
    origin at ${\bf x}=0$.  
    We  choose  a function $d(Z)$ as follows.
The constraint, ${\cal R}= -12/\ell^2$, after linearizing
 in $d$ and its derivatives, has the solution $d(Z) =d_0 + c_0 Z^4$, where
$c_0$ and $d_0$ are constants.  
For the case $c_0=0$,     (\ref{Psi-1})   is an exact solution
and  the   RS1 limit ($\Omega_2\rightarrow 0$) is easily taken.
We will work in RS2 and take  $c_0$  as a small nonzero parameter.

    On each slice $Z$=constant, we  now transform to 
  spherical coordinates centered at ${\bf x}=0$, with $z=\rho \cos\theta$,
and
 expand $\Psi$  in    Legendre polynomials $P_{k}(\cos\theta)$  for $\rho > d$,
 \begin{equation}
  \label{Psi-2}
 \Psi =1 + \frac{\rho_0}{\rho} +  
\frac{\rho_0}{\rho}  \sum_{j=1}^\infty \left(\frac{d}{\rho}\right)^{2j} P_{2j}(\cos\theta) \ .
\end{equation} 
 On each slice $Z$=constant,  we  will  take   $\rho_0 \gg d$, which
  describes a
 3-dimensional   black hole    \cite{Brill-Lindquist}, and   $d$  parametrizes  
   the    2-dimensional apparent horizon's distortion    from the sphere $\rho=\rho_0$.
The  full   3-dimensional apparent horizon   
       (the union of   the 2-dimensional 
 apparent horizons) therefore
   describes a  distorted black string.

 On each slice $Z$=constant,  as in  \cite{Brill-Lindquist}, 
the  surface $\rho(\theta)$  of the 2-dimensional    apparent horizon    can be found as the sum of  Legendre polynomials  
   that 
  minimizes the area $A_2$,
 \begin{equation}
A_2 = 2\pi    
\int_0^{\pi}  d\theta\, \Psi^4 \rho \sqrt{\rho^2 + \left( \frac{d\rho}{d\theta}\right)^2} \ .
 \end{equation}
To lowest order in  $(d/\rho_0) \ll 1$,   we   find the  apparent horizon
on each slice $Z$=constant is
  \begin{equation}
  \label{bs-AH}
 \rho(\theta) = \rho_0
 \left[ 1  +  \frac{5}{7}\left(\frac{d}{\rho_0} \right)^2 P_2(\cos\theta) \right] \ .
 \end{equation}
 This agrees  with the numerical results of \cite{Brill-Lindquist},
and gives, to lowest order  in $(d/\rho_0$),
\begin{equation}
\label{bs-area-2D}
A_2(Z)  =  
64\pi \rho_0^2 
\left[
1 - \frac{5 }{7}\left(\frac{d}{\rho_0}\right)^4
\right] \ .
 \end{equation}
 Integrating in  $Z$ gives
the 3-dimensional
 area of the   full  apparent horizon in the   black string geometry,     
\begin{equation}
 \label{bs-area}
A =      \int_{Z_1}^{Z_2} dZ\, \Omega^3   
A_2  
=
32\pi  w 
\left(
\rho_0^2  - \frac{5Q}{7\rho_0^2}  
\right) \ ,
 \end{equation}
where, with   $\Omega_i$   the warp factor  at each brane,
  \begin{equation}
 w = \ell \left(\Omega_1^2 - \Omega_2^2\right)
 \ , 
\quad
Q =  d_0^4 
+
\frac{8 \, c_0 \, d_0^3 \, \ell^4}{\Omega_1\Omega_2(\Omega_1+\Omega_2)} \ .
 \end{equation}
We defined the 
 mass $M$   in  \cite{paper-1-first-law}, which for
(\ref{Psi-2})    gives
  \begin{equation}
  \label{bs-mass}
 G_5 M =w  \rho_0 \ .
 \end{equation}
To lowest order in $Q$, combining  (\ref{bs-area})\textendash(\ref{bs-mass})  gives
\begin{equation}
\label{bs mass function}
\left(\frac{G_5 M}{w}\right)^2 =  \left(\frac{A}{32\pi w } \right)+ \frac{5}{7} \left(\frac{32\pi w }{A} \right)  Q \ .
 \end{equation}
We   now     apply our variational principle: we extremize 
  $M$  in (\ref{bs mass function}) at fixed  $A$, $\ell$, and $\Omega_i$. 
  This yields the conditions $c_0=0$ and $d_0=0$, which   we conclude   
  describes  a static black string. 
  We can verify this    directly,
 since  $d=0$ in  (\ref{Psi-2}) gives
   $\Psi = 1+\rho_0/\rho$ and the apparent horizon 
 (\ref{bs-AH})
 is located at $\rho=\rho_0$.  This is indeed the initial geometry
of  the static black string  \cite{black-string}
 in isotropic coordinates.

 We can also deduce that 
 the   evolution of  the distorted black string, with $d \neq 0$, will not be static, since    
 each slice $Z$=constant is   the initial geometry for an attracting
  two-body problem  \cite{Brill-Lindquist}, and
 the 2-dimensional apparent horizon   considered above    
  describes a black hole formed by, and surrounding,
  two closely separated smaller black holes (with 
a small minimal surface  surrounding each   point  ${\bf x}=\pm{\bf x}_0$).
From the perspective in each slice $Z$=constant,
as in    \cite{Brill-Lindquist}, the
 two small interior black holes 
  will coalesce   as
  the initial data evolves, due to   
  mutual gravitational attraction.  This
     results in a time-dependent geometry on each slice $Z$=constant,
 and   results in    a time-dependent black string 
      geometry in the bulk perspective.
  
\section{Discussion \label{conclusion}}

  The variational principle   developed in this paper states that
  for an asymptotically RS black hole
  initially at rest,   initial data that extremizes the mass yields a static black hole,
   for variations at fixed values of the apparent horizon area and the 
   remaining physical variables in the first law.  
 It would be interesting to investigate the consequences of holding fewer variables  fixed.
An example of this  in 4-dimensional spacetime
 is Hawking's proof     \cite{hawking}    that the
  static (Schwarzschild) black hole is an extremum of mass   
at fixed apparent horizon area but   arbitrary angular momentum. 
 
Our example application of the   variational principle   
 to a trial solution serves as
a prelude to the approach we will take in the next paper \cite{paper-3-extrema} in this series.
In \cite{paper-3-extrema}, we
will  
 conclude that solutions exist for small static black holes in RS2, 
 both on and off the brane, as special members of 
 a general family of initially static black holes.
This family of black hole initial data will also indicate   
  that a   small black hole  on an orbifold-symmetric brane in RS2   is stable against leaving the brane, 
which generalizes 
  to other  orbifold-symmetric braneworld models, 
and is   an important result for future collider experiments.

\begin{acknowledgments}
   This research was supported in part by the National Science Foundation
   under Grant No.\ NSF PHY11-25915.
\end{acknowledgments}


\bibliography{var-principle-v2}

\begin{thebibliography}{24}%
\makeatletter
\providecommand \@ifxundefined [1]{%
 \@ifx{#1\undefined}
}%
\providecommand \@ifnum [1]{%
 \ifnum #1\expandafter \@firstoftwo
 \else \expandafter \@secondoftwo
 \fi
}%
\providecommand \@ifx [1]{%
 \ifx #1\expandafter \@firstoftwo
 \else \expandafter \@secondoftwo
 \fi
}%
\providecommand \natexlab [1]{#1}%
\providecommand \enquote  [1]{``#1''}%
\providecommand \bibnamefont  [1]{#1}%
\providecommand \bibfnamefont [1]{#1}%
\providecommand \citenamefont [1]{#1}%
\providecommand \href@noop [0]{\@secondoftwo}%
\providecommand \href [0]{\begingroup \@sanitize@url \@href}%
\providecommand \@href[1]{\@@startlink{#1}\@@href}%
\providecommand \@@href[1]{\endgroup#1\@@endlink}%
\providecommand \@sanitize@url [0]{\catcode `\\12\catcode `\$12\catcode
  `\&12\catcode `\#12\catcode `\^12\catcode `\_12\catcode `\%12\relax}%
\providecommand \@@startlink[1]{}%
\providecommand \@@endlink[0]{}%
\providecommand \url  [0]{\begingroup\@sanitize@url \@url }%
\providecommand \@url [1]{\endgroup\@href {#1}{\urlprefix }}%
\providecommand \urlprefix  [0]{URL }%
\providecommand \Eprint [0]{\href }%
\providecommand \doibase [0]{http://dx.doi.org/}%
\providecommand \selectlanguage [0]{\@gobble}%
\providecommand \bibinfo  [0]{\@secondoftwo}%
\providecommand \bibfield  [0]{\@secondoftwo}%
\providecommand \translation [1]{[#1]}%
\providecommand \BibitemOpen [0]{}%
\providecommand \bibitemStop [0]{}%
\providecommand \bibitemNoStop [0]{.\EOS\space}%
\providecommand \EOS [0]{\spacefactor3000\relax}%
\providecommand \BibitemShut  [1]{\csname bibitem#1\endcsname}%
\let\auto@bib@innerbib\@empty
\bibitem [{\citenamefont {Hawking}(1973)}]{hawking}%
  \BibitemOpen
  \bibfield  {author} {\bibinfo {author} {\bibfnamefont {S.~W.}\ \bibnamefont
  {Hawking}},\ }\href {\doibase 10.1007/BF01646744} {\bibfield  {journal}
  {\bibinfo  {journal} {Commun. Math. Phys.}\ }\textbf {\bibinfo {volume}
  {33}},\ \bibinfo {pages} {323} (\bibinfo {year} {1973})}\BibitemShut
  {NoStop}%
\bibitem [{\citenamefont {Sudarsky}\ and\ \citenamefont
  {Wald}(1992)}]{sudarsky-wald}%
  \BibitemOpen
  \bibfield  {author} {\bibinfo {author} {\bibfnamefont {D.}~\bibnamefont
  {Sudarsky}}\ and\ \bibinfo {author} {\bibfnamefont {R.~M.}\ \bibnamefont
  {Wald}},\ }\href {\doibase 10.1103/PhysRevD.46.1453} {\bibfield  {journal}
  {\bibinfo  {journal} {Phys. Rev.}\ }\textbf {\bibinfo {volume} {D46}},\
  \bibinfo {pages} {1453} (\bibinfo {year} {1992})}\BibitemShut {NoStop}%
\bibitem [{\citenamefont {Chrusciel}\ and\ \citenamefont
  {Wald}(1994)}]{chrusciel-wald}%
  \BibitemOpen
  \bibfield  {author} {\bibinfo {author} {\bibfnamefont {P.~T.}\ \bibnamefont
  {Chrusciel}}\ and\ \bibinfo {author} {\bibfnamefont {R.~M.}\ \bibnamefont
  {Wald}},\ }\href {\doibase 10.1007/BF02101463} {\bibfield  {journal}
  {\bibinfo  {journal} {Commun.Math.Phys.}\ }\textbf {\bibinfo {volume}
  {163}},\ \bibinfo {pages} {561} (\bibinfo {year} {1994})},\ \Eprint
  {http://arxiv.org/abs/gr-qc/9304009} {arXiv:gr-qc/9304009} \BibitemShut
  {NoStop}%
\bibitem [{\citenamefont {Randall}\ and\ \citenamefont
  {Sundrum}(1999{\natexlab{a}})}]{RS1}%
  \BibitemOpen
  \bibfield  {author} {\bibinfo {author} {\bibfnamefont {L.}~\bibnamefont
  {Randall}}\ and\ \bibinfo {author} {\bibfnamefont {R.}~\bibnamefont
  {Sundrum}},\ }\href {\doibase 10.1103/PhysRevLett.83.3370} {\bibfield
  {journal} {\bibinfo  {journal} {Phys. Rev. Lett.}\ }\textbf {\bibinfo
  {volume} {83}},\ \bibinfo {pages} {3370} (\bibinfo {year}
  {1999}{\natexlab{a}})},\ \Eprint {http://arxiv.org/abs/hep-ph/9905221}
  {arXiv:hep-ph/9905221} \BibitemShut {NoStop}%
\bibitem [{\citenamefont {Randall}\ and\ \citenamefont
  {Sundrum}(1999{\natexlab{b}})}]{RS2}%
  \BibitemOpen
  \bibfield  {author} {\bibinfo {author} {\bibfnamefont {L.}~\bibnamefont
  {Randall}}\ and\ \bibinfo {author} {\bibfnamefont {R.}~\bibnamefont
  {Sundrum}},\ }\href {\doibase 10.1103/PhysRevLett.83.4690} {\bibfield
  {journal} {\bibinfo  {journal} {Phys. Rev. Lett.}\ }\textbf {\bibinfo
  {volume} {83}},\ \bibinfo {pages} {4690} (\bibinfo {year}
  {1999}{\natexlab{b}})},\ \Eprint {http://arxiv.org/abs/hep-th/9906064}
  {arXiv:hep-th/9906064} \BibitemShut {NoStop}%
\bibitem [{\citenamefont {Arkani-Hamed}\ \emph {et~al.}(2001)\citenamefont
  {Arkani-Hamed}, \citenamefont {Porrati},\ and\ \citenamefont
  {Randall}}]{RS-AdS-CFT}%
  \BibitemOpen
  \bibfield  {author} {\bibinfo {author} {\bibfnamefont {N.}~\bibnamefont
  {Arkani-Hamed}}, \bibinfo {author} {\bibfnamefont {M.}~\bibnamefont
  {Porrati}}, \ and\ \bibinfo {author} {\bibfnamefont {L.}~\bibnamefont
  {Randall}},\ }\href {\doibase 10.1088/1126-6708/2001/08/017} {\bibfield
  {journal} {\bibinfo  {journal} {JHEP}\ }\textbf {\bibinfo {volume} {0108}},\
  \bibinfo {pages} {017} (\bibinfo {year} {2001})},\ \Eprint
  {http://arxiv.org/abs/hep-th/0012148} {arXiv:hep-th/0012148} \BibitemShut
  {NoStop}%
\bibitem [{\citenamefont {Banks}\ and\ \citenamefont
  {Fischler}(1999)}]{Banks-Fischler}%
  \BibitemOpen
  \bibfield  {author} {\bibinfo {author} {\bibfnamefont {T.}~\bibnamefont
  {Banks}}\ and\ \bibinfo {author} {\bibfnamefont {W.}~\bibnamefont
  {Fischler}},\ }\href@noop {} {\  (\bibinfo {year} {1999})},\ \Eprint
  {http://arxiv.org/abs/hep-th/9906038} {arXiv:hep-th/9906038} \BibitemShut
  {NoStop}%
\bibitem [{\citenamefont {Giddings}\ and\ \citenamefont
  {Thomas}(2002)}]{Giddings}%
  \BibitemOpen
  \bibfield  {author} {\bibinfo {author} {\bibfnamefont {S.~B.}\ \bibnamefont
  {Giddings}}\ and\ \bibinfo {author} {\bibfnamefont {S.~D.}\ \bibnamefont
  {Thomas}},\ }\href {\doibase 10.1103/PhysRevD.65.056010} {\bibfield
  {journal} {\bibinfo  {journal} {Phys. Rev.}\ }\textbf {\bibinfo {volume}
  {D65}},\ \bibinfo {pages} {056010} (\bibinfo {year} {2002})},\ \Eprint
  {http://arxiv.org/abs/hep-ph/0106219} {arXiv:hep-ph/0106219} \BibitemShut
  {NoStop}%
\bibitem [{\citenamefont {Chatrchyan}\ \emph {et~al.}(2012)\citenamefont
  {Chatrchyan} \emph {et~al.}}]{LHC-RS-extra-dimensions}%
  \BibitemOpen
  \bibfield  {author} {\bibinfo {author} {\bibfnamefont {S.}~\bibnamefont
  {Chatrchyan}} \emph {et~al.} (\bibinfo {collaboration} {CMS Collaboration}),\
  }\href {\doibase 10.1103/PhysRevLett.108.111801} {\bibfield  {journal}
  {\bibinfo  {journal} {Phys. Rev. Lett.}\ }\textbf {\bibinfo {volume} {108}},\
  \bibinfo {pages} {111801} (\bibinfo {year} {2012})},\ \Eprint
  {http://arxiv.org/abs/1112.0688} {arXiv:1112.0688} \BibitemShut {NoStop}%
\bibitem [{\citenamefont {Chatrchyan}\ \emph {et~al.}(2013)\citenamefont
  {Chatrchyan} \emph {et~al.}}]{LHC-RS-black-holes}%
  \BibitemOpen
  \bibfield  {author} {\bibinfo {author} {\bibfnamefont {S.}~\bibnamefont
  {Chatrchyan}} \emph {et~al.} (\bibinfo {collaboration} {CMS Collaboration}),\
  }\href {\doibase 10.1007/JHEP07(2013)178} {\bibfield  {journal} {\bibinfo
  {journal} {JHEP}\ }\textbf {\bibinfo {volume} {1307}},\ \bibinfo {pages}
  {178} (\bibinfo {year} {2013})},\ \Eprint {http://arxiv.org/abs/1303.5338}
  {arXiv:1303.5338} \BibitemShut {NoStop}%
\bibitem [{\citenamefont {Tanaka}\ and\ \citenamefont {Montes}(2000)}]{Tanaka}%
  \BibitemOpen
  \bibfield  {author} {\bibinfo {author} {\bibfnamefont {T.}~\bibnamefont
  {Tanaka}}\ and\ \bibinfo {author} {\bibfnamefont {X.}~\bibnamefont
  {Montes}},\ }\href {\doibase 10.1016/S0550-3213(00)00328-X} {\bibfield
  {journal} {\bibinfo  {journal} {Nucl. Phys.}\ }\textbf {\bibinfo {volume}
  {B582}},\ \bibinfo {pages} {259} (\bibinfo {year} {2000})},\ \Eprint
  {http://arxiv.org/abs/hep-th/0001092} {arXiv:hep-th/0001092} \BibitemShut
  {NoStop}%
\bibitem [{\citenamefont {Kudoh}\ \emph {et~al.}(2003)\citenamefont {Kudoh},
  \citenamefont {Tanaka},\ and\ \citenamefont {Nakamura}}]{Kudoh-smallBH-1}%
  \BibitemOpen
  \bibfield  {author} {\bibinfo {author} {\bibfnamefont {H.}~\bibnamefont
  {Kudoh}}, \bibinfo {author} {\bibfnamefont {T.}~\bibnamefont {Tanaka}}, \
  and\ \bibinfo {author} {\bibfnamefont {T.}~\bibnamefont {Nakamura}},\ }\href
  {\doibase 10.1103/PhysRevD.68.024035} {\bibfield  {journal} {\bibinfo
  {journal} {Phys. Rev.}\ }\textbf {\bibinfo {volume} {D68}},\ \bibinfo {pages}
  {024035} (\bibinfo {year} {2003})},\ \Eprint
  {http://arxiv.org/abs/gr-qc/0301089} {arXiv:gr-qc/0301089} \BibitemShut
  {NoStop}%
\bibitem [{\citenamefont {Kudoh}(2004{\natexlab{a}})}]{Kudoh-smallBH-2}%
  \BibitemOpen
  \bibfield  {author} {\bibinfo {author} {\bibfnamefont {H.}~\bibnamefont
  {Kudoh}},\ }\href {\doibase 10.1143/PTP.110.1059} {\bibfield  {journal}
  {\bibinfo  {journal} {Prog. Theor. Phys.}\ }\textbf {\bibinfo {volume}
  {110}},\ \bibinfo {pages} {1059} (\bibinfo {year} {2004}{\natexlab{a}})},\
  \Eprint {http://arxiv.org/abs/hep-th/0306067} {arXiv:hep-th/0306067}
  \BibitemShut {NoStop}%
\bibitem [{\citenamefont {Kudoh}(2004{\natexlab{b}})}]{Kudoh-smallBH-6D}%
  \BibitemOpen
  \bibfield  {author} {\bibinfo {author} {\bibfnamefont {H.}~\bibnamefont
  {Kudoh}},\ }\href {\doibase 10.1103/PhysRevD.69.104019} {\bibfield  {journal}
  {\bibinfo  {journal} {Phys. Rev.}\ }\textbf {\bibinfo {volume} {D69}},\
  \bibinfo {pages} {104019} (\bibinfo {year} {2004}{\natexlab{b}})},\ \Eprint
  {http://arxiv.org/abs/hep-th/0401229} {arXiv:hep-th/0401229} \BibitemShut
  {NoStop}%
\bibitem [{\citenamefont {Figueras}\ and\ \citenamefont
  {Wiseman}(2011)}]{Figueras-Wiseman}%
  \BibitemOpen
  \bibfield  {author} {\bibinfo {author} {\bibfnamefont {P.}~\bibnamefont
  {Figueras}}\ and\ \bibinfo {author} {\bibfnamefont {T.}~\bibnamefont
  {Wiseman}},\ }\href {\doibase 10.1103/PhysRevLett.107.081101} {\bibfield
  {journal} {\bibinfo  {journal} {Phys. Rev. Lett.}\ }\textbf {\bibinfo
  {volume} {107}},\ \bibinfo {pages} {081101} (\bibinfo {year} {2011})},\
  \Eprint {http://arxiv.org/abs/1105.2558} {arXiv:1105.2558} \BibitemShut
  {NoStop}%
\bibitem [{\citenamefont {Abdolrahimi}\ \emph {et~al.}(2013)\citenamefont
  {Abdolrahimi}, \citenamefont {Cattoen}, \citenamefont {Page},\ and\
  \citenamefont {Yaghoobpour-Tari}}]{Abdolrahimi}%
  \BibitemOpen
  \bibfield  {author} {\bibinfo {author} {\bibfnamefont {S.}~\bibnamefont
  {Abdolrahimi}}, \bibinfo {author} {\bibfnamefont {C.}~\bibnamefont
  {Cattoen}}, \bibinfo {author} {\bibfnamefont {D.~N.}\ \bibnamefont {Page}}, \
  and\ \bibinfo {author} {\bibfnamefont {S.}~\bibnamefont {Yaghoobpour-Tari}},\
  }\href {\doibase 10.1016/j.physletb.2013.02.034} {\bibfield  {journal}
  {\bibinfo  {journal} {Phys. Lett.}\ }\textbf {\bibinfo {volume} {B720}},\
  \bibinfo {pages} {405} (\bibinfo {year} {2013})},\ \Eprint
  {http://arxiv.org/abs/1206.0708} {arXiv:1206.0708} \BibitemShut {NoStop}%
\bibitem [{\citenamefont {Emparan}\ \emph
  {et~al.}(2000{\natexlab{a}})\citenamefont {Emparan}, \citenamefont
  {Horowitz},\ and\ \citenamefont {Myers}}]{ehm}%
  \BibitemOpen
  \bibfield  {author} {\bibinfo {author} {\bibfnamefont {R.}~\bibnamefont
  {Emparan}}, \bibinfo {author} {\bibfnamefont {G.~T.}\ \bibnamefont
  {Horowitz}}, \ and\ \bibinfo {author} {\bibfnamefont {R.~C.}\ \bibnamefont
  {Myers}},\ }\href {\doibase 10.1088/1126-6708/2000/01/007} {\bibfield
  {journal} {\bibinfo  {journal} {JHEP}\ }\textbf {\bibinfo {volume} {0001}},\
  \bibinfo {pages} {007} (\bibinfo {year} {2000}{\natexlab{a}})},\ \Eprint
  {http://arxiv.org/abs/hep-th/9911043} {arXiv:hep-th/9911043} \BibitemShut
  {NoStop}%
\bibitem [{\citenamefont {Emparan}\ \emph
  {et~al.}(2000{\natexlab{b}})\citenamefont {Emparan}, \citenamefont
  {Horowitz},\ and\ \citenamefont {Myers}}]{ehm-2}%
  \BibitemOpen
  \bibfield  {author} {\bibinfo {author} {\bibfnamefont {R.}~\bibnamefont
  {Emparan}}, \bibinfo {author} {\bibfnamefont {G.~T.}\ \bibnamefont
  {Horowitz}}, \ and\ \bibinfo {author} {\bibfnamefont {R.~C.}\ \bibnamefont
  {Myers}},\ }\href {\doibase 10.1088/1126-6708/2000/01/021} {\bibfield
  {journal} {\bibinfo  {journal} {JHEP}\ }\textbf {\bibinfo {volume} {0001}},\
  \bibinfo {pages} {021} (\bibinfo {year} {2000}{\natexlab{b}})},\ \Eprint
  {http://arxiv.org/abs/hep-th/9912135} {arXiv:hep-th/9912135} \BibitemShut
  {NoStop}%
\bibitem [{\citenamefont {Fraser}\ and\ \citenamefont
  {Eardley}(2014{\natexlab{a}})}]{paper-1-first-law}%
  \BibitemOpen
  \bibfield  {author} {\bibinfo {author} {\bibfnamefont {S.}~\bibnamefont
  {Fraser}}\ and\ \bibinfo {author} {\bibfnamefont {D.~M.}\ \bibnamefont
  {Eardley}},\ }\href@noop {} {\  (\bibinfo {year} {2014}{\natexlab{a}})},\
  \Eprint {http://arxiv.org/abs/1408.4425} {arXiv:1408.4425} \BibitemShut
  {NoStop}%
\bibitem [{\citenamefont {Fraser}\ and\ \citenamefont
  {Eardley}(2014{\natexlab{b}})}]{paper-3-extrema}%
  \BibitemOpen
  \bibfield  {author} {\bibinfo {author} {\bibfnamefont {S.}~\bibnamefont
  {Fraser}}\ and\ \bibinfo {author} {\bibfnamefont {D.~M.}\ \bibnamefont
  {Eardley}},\ }\href@noop {} {\  (\bibinfo {year} {2014}{\natexlab{b}})},\
  \Eprint {http://arxiv.org/abs/1409.0884} {arXiv:1409.0884} \BibitemShut
  {NoStop}%
\bibitem [{\citenamefont {Israel}(1966)}]{israel}%
  \BibitemOpen
  \bibfield  {author} {\bibinfo {author} {\bibfnamefont {W.}~\bibnamefont
  {Israel}},\ }\href {\doibase 10.1007/BF02730328} {\bibfield  {journal}
  {\bibinfo  {journal} {Nuovo Cim.}\ }\textbf {\bibinfo {volume} {B44S10}},\
  \bibinfo {pages} {1} (\bibinfo {year} {1966})}\BibitemShut {NoStop}%
\bibitem [{\citenamefont {Chamblin}\ \emph {et~al.}(2000)\citenamefont
  {Chamblin}, \citenamefont {Hawking},\ and\ \citenamefont
  {Reall}}]{black-string}%
  \BibitemOpen
  \bibfield  {author} {\bibinfo {author} {\bibfnamefont {A.}~\bibnamefont
  {Chamblin}}, \bibinfo {author} {\bibfnamefont {S.}~\bibnamefont {Hawking}}, \
  and\ \bibinfo {author} {\bibfnamefont {H.}~\bibnamefont {Reall}},\ }\href
  {\doibase 10.1103/PhysRevD.61.065007} {\bibfield  {journal} {\bibinfo
  {journal} {Phys. Rev.}\ }\textbf {\bibinfo {volume} {D61}},\ \bibinfo {pages}
  {065007} (\bibinfo {year} {2000})},\ \Eprint
  {http://arxiv.org/abs/hep-th/9909205} {arXiv:hep-th/9909205} \BibitemShut
  {NoStop}%
\bibitem [{\citenamefont {Daners}(2009)}]{inverse-positivity}%
  \BibitemOpen
  \bibfield  {author} {\bibinfo {author} {\bibfnamefont {D.}~\bibnamefont
  {Daners}},\ }\href {\doibase 10.1007/s00013-008-2918-z} {\bibfield  {journal}
  {\bibinfo  {journal} {Arch. Math.}\ }\textbf {\bibinfo {volume} {92}},\
  \bibinfo {pages} {57} (\bibinfo {year} {2009})}\BibitemShut {NoStop}%
\bibitem [{\citenamefont {Brill}\ and\ \citenamefont
  {Lindquist}(1963)}]{Brill-Lindquist}%
  \BibitemOpen
  \bibfield  {author} {\bibinfo {author} {\bibfnamefont {D.~R.}\ \bibnamefont
  {Brill}}\ and\ \bibinfo {author} {\bibfnamefont {R.~W.}\ \bibnamefont
  {Lindquist}},\ }\href {\doibase 10.1103/PhysRev.131.471} {\bibfield
  {journal} {\bibinfo  {journal} {Phys. Rev.}\ }\textbf {\bibinfo {volume}
  {131}},\ \bibinfo {pages} {471} (\bibinfo {year} {1963})}\BibitemShut
  {NoStop}%
\end{thebibliography}%

\end{document}